\documentclass[12pt,a4paper]{article}
\usepackage[utf8]{inputenc}
\usepackage[english]{babel}
\usepackage{amsmath}
\numberwithin{equation}{section}
\usepackage{booktabs}
\usepackage{amsfonts}
\usepackage{amssymb}
\usepackage{mathtools}
\usepackage{color}
\usepackage{graphicx}
\usepackage{changes}
\usepackage{caption}
\usepackage{subcaption}
\usepackage{authblk}
\usepackage[left=1.8cm,right=1.8cm,top=2cm,bottom=2cm]{geometry}
\def\ev{\hat{\mathbf{e}}}
\def\betac{{\beta_{\rm cr}}}
\def\muext{{\mu_{\rm ext}}}
\title{Circumferential buckling  of a hydrogel tube emptying upon dehydration}
\author[1]{M. Curatolo}
\author[2]{F. Lisi}
\author[2]{G. Napoli}
\author[3]{P. Nardinocchi}
\affil[1]{\small Dipartimento di Architettura, Università di Roma Tre, Roma, Italy}
\affil[2]{\small Dipartimento di Matematica e Fisica ‘E. De Giorgi’, Università del Salento, Lecce, Italy}
\affil[3]{\small Dipartimento di Ingegneria Strutturale e Geotecnica, Sapienza Università di Roma, Italy}
\date{}
\begin{document}
\maketitle
\begin{abstract}
A cylindrical hydrogel tube, completely submerged in water, hydrates by swelling and filling its internal cavity. When it comes back into contact with air, it dehydrates: the tube thus expels the solvent through the walls, shrinking. This dehydration process causes a depression in the tube cavity, which can lead to circumferential buckling. Here we study the occurrence of such buckling using a continuous model that combines non-linear elasticity with Flory-Rehner theory, to take into account both the large deformations and the active behavior of the hydrogel. In quasi-static approximation, we use the incremental deformation formalism, extended to the chemo-mechanical equations, to determine the threshold value of the enclosed volume at which buckling is triggered. This critical value is found to depend on the shell thickness, chemical potential and constitutive features. The results obtained are in good agreement with the results of the finite element simulations of the complete dynamic problem.
\end{abstract}

\section{Introduction}
Hydrogels are colloidal gels consisting of a self-supporting polymer network in which water is the dispersed medium. In recent years, they have been extensively studied because they can undergo large deformations, actively swelling and shrinking as a result of absorbing and releasing solvents in response to specific environmental conditions such as humidity, temperature and pH.  
Hydrogel-like materials are common in nature: the release of solvent in response to specific stimuli is used to fulfil precise functional requirements, such as the onset of specific deformation patterns and the distribution of fixed amounts of solvent to the external environment \cite{dawson1997,burgert2009,noblin2012,erb2013}. In industrial processes, hydrogels have also gained much attention for their promising role in a wide range of applications in recent emerging technologies, such as microfluidics, 3D bioprinting technology and drug delivery \cite{burgert2009,erb2013,llorens2016,egunov2016,SHI2016,li2016,GOY2019,el2013}. For further details about synthesis, properties, and applications of hydrogels we suggest the reading of S. B. Majee's book \cite{Majee_16} and E. M. Ahmed's review \cite{AHMED2015}. 

In this paper we study a two-dimensional problem related to the cross-sectional instability that occurs in a cylindrical hydrogel tube that passes from a completely wet state, with the cavity filled with water, to a state where the outer wall is exposed to air. The depression of the cavity, which is a consequence of water draining through the walls, causes the buckling from circular to wavy shapes of the tube cross-section. 

The circumferential buckling is a common pattern of  pressurized elastic rings   \cite{Carrier:1947,Dion:1995}, bi-layer tubes \cite{Emuna:2020} and growing hollow cylinders \cite{moulton2011}. A cylindrical elastic tube under a uniform radial external pressure (or, equivalently, under an inner depression)  might, indeed,  buckle circumferentially to a non-circular cross-section at a critical threshold. A number of studies have been performed in the literature to inspect how mechanical behaviours and geometrical properties of the tube affect the circumferential buckling, in terms of the preferred mode number and the critical threshold at which it occurs (see \cite{moulton2011} and references therein). In bi-layer pipes, instead, the circumferential wrinkling may be induced by the radial incompatibility between the layers \cite{Emuna:2020}.  
Other works \cite{moulton2011, Cao2012,jia2015, Jin:2018, Liu:2021} analyse this buckling instability in growing elastic tubes. Therein, growth is defined with respect to the initial reference configuration and  the cumulative effect of the incremental growth process is represented by the growth tensor. Buckling are then parameterized in term of radial and circumferential growth. Thus, it turns out a change in thickness due to growth induces a dramatic impact on circumferential buckling, both in the critical pressure and the buckling pattern.

A hollow cylinder-shaped hydrogel shell, completely immersed in water, hydrates by thickening and filling the internal cavity. When brought back into contact with air it dehydrates: the shell expels the solvent through the walls, becoming thinner. The flow of solvent through the walls is induced by the difference in chemical potential between the inner and outer walls \cite{curatolo2018, curatolo2020}. Assuming that the inner cavity is always occupied by solvent, dehydration causes a depression in the cavity that can lead to circumferential buckling.
As already observed in a similar study on spherical capsules \cite{curatolo2021}, and   in contrast to problems of pressure or growth-induced instability, the analysis here requires some extra care: the classical control parameter of the instability, which is the pressure exerted on the internal walls, is not known {\it a priori}. 
Indeed, in this problem, cavity pressure is the result of a complex interplay between elasticity, geometry and solvent flow induced by the difference in chemical potential between the inner and outer walls. Thus chemical potential, which drives the dehydration process, is the actual control parameter.

The article is organised as follows. In Section 2 we present the physical-mathematical model underlying the swelling of hydrogels, which combines elasticity theory with Flory-Renher theory. In Section 3, we report on the analysis of the bifurcation from the axisymmetric to the circumferentially wavy solution, under quasi-static hypothesis and using the method of incremental equations. We thus obtain a linear boundary problem that has to be simultaneously solved with the axisymmetric problem in finite deformations. As a result of this semi-analytical method, we obtain the critical threshold and the corresponding buckling mode, depending on the geometrical and constitutive parameters of the model. In Section 4, we discuss the results and compare them to the finite element simulations of the nonlinear dynamic model. Finally, we draw our conclusions in Section 5. 

\section{Model}
The hydrogel model is set within the framework of stress-diffusion theories, which view the liquid-polymer mixture as a single homogenized continuum body, allow for a mass flux of the solvent, inherit  the balance equations of the theory from the principle of null working and the solvent mass conservation law and include admissible constitutive processes which are consistent with the thermodynamic principles \cite{lucantonio2013}.

\begin{figure}[t]
\centering
\includegraphics[scale=1]{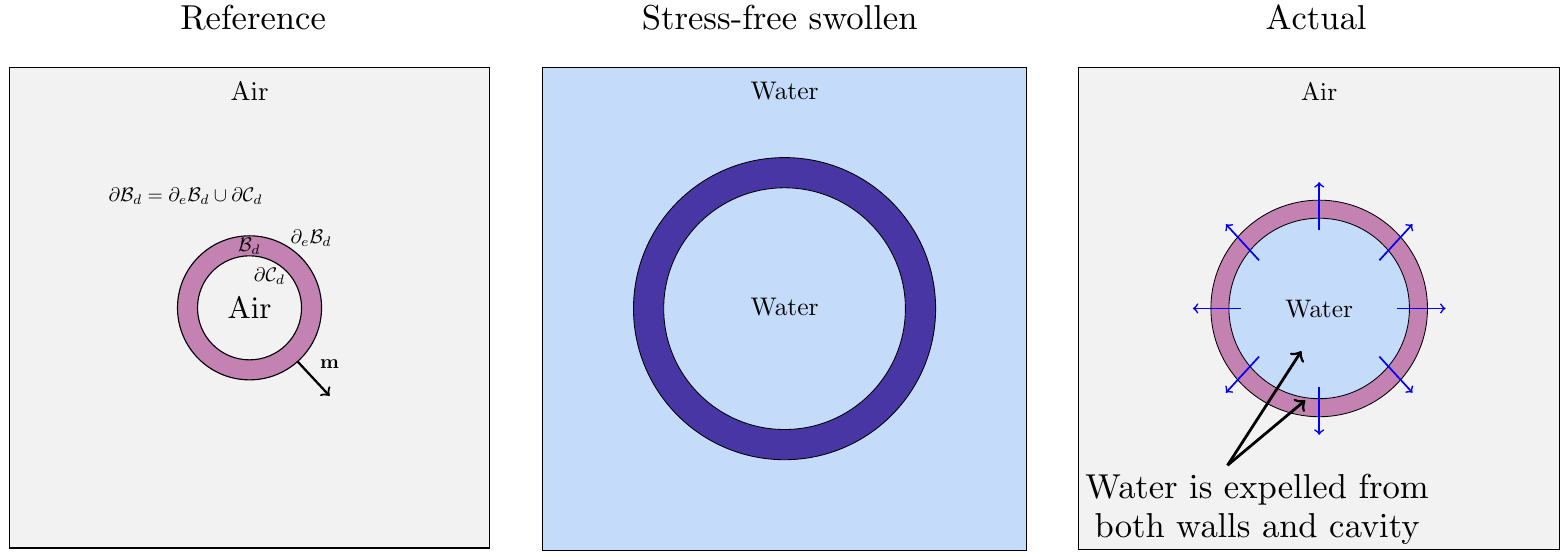}
\caption{Schematic representation of the hydration/dehydration process. Dry state of the tube (left). Initial steady stress-free swollen state of the tube: water fills the cavity and the external environment (middle). After exposure to air, the dehydration process starts and water is expelled from both the walls and the tube (right)}
\label{schematic}
\end{figure}

The dry state $\mathcal{B}_d$ of the gel is used as reference configuration (see Figure \ref{schematic}). The chemo-mechanical description of the body is, then, complete once both the displacement field $\textbf{u}_d$, which gives the actual position $\mathbf{x}=\mathbf{X}_d+\mathbf{u}_d(t,\mathbf{X}_d)$ of a point $\mathbf{X}_d\in\mathcal{B}_d$ at time $t$, and the water concentration $c_d(t,\mathbf{X}_d)$, that gives the moles of solvent per unit dry volume at $\mathbf{x}$, are known.

The free energy $\psi_d$ per unit dry volume  of the gel depends on the neo-Hookean elastic behaviour of the shell and on the polymer-water interplay as prescribed by the Flory-Rehner thermodynamic model \cite{flory1943I, flory1943II}. Furthermore, we assume that any change in volume of the body is accompanied by an equivalent uptake or release of water content, that is
\begin{equation}
\label{vc}
\det\mathbf{F}_d=1+\Omega c_d\eqqcolon J_d\,,
\end{equation}
where $\Omega$ is the molar volume of the water and $\det\mathbf{F}_d$ the determinant of the deformation gradient in a plane strain framework. Thus, the free energy density $\psi_d^{(r)}$ can be written as
\begin{equation}
\psi_d^{(r)}(\mathbf{F}_d,c_d,p)\coloneqq\frac{G_d}{2}\big(\mathbf{F}_d\cdot \mathbf{F}_d-3\big)+\frac{\mathcal{R}T}{\Omega}h(c_d)-p\big(\det\mathbf{F}_d-J_d\big)\,,
\end{equation}
with
\begin{equation}
h(c_d)\coloneqq\Omega c_d\log\frac{\Omega c_d}{1+\Omega c_d}+\chi\frac{\Omega c_d}{1+\Omega c_d}\,,
\end{equation}
and the shear-modulus $G_d$  of the shell, the Flory parameter  $\chi$, the absolute temperature $T$ and  the gas constant $\mathcal{R}$, assumed to be known. The field $p$ represents the Lagrange multiplier related to the mechanical incompressibility constraint \eqref{vc}.

The reference stress tensor $\mathbf{S}_d$ and the chemical potential $\mu$ of the liquid within the gel are
\begin{equation}
\label{oggetti}
\begin{aligned}
\mathbf{S}_d&\coloneqq\frac{\partial\psi_d^{(r)}}{\partial\mathbf{F}_d}=G_d\textbf{F}_d-pJ_d\mathbf{F}_d^{-T}=\widehat{\mathbf{S}_d}(\mathbf{F}_d)-pJ_d\mathbf{F}_d^{-T}\,,\\
\mu &\coloneqq\mu_{w}+\frac{\partial\psi_d^{(r)}}{\partial c_d}=\mu_{w}+\mathcal{R}T\left(\log\frac{J_d-1}{J_d}+\frac{1}{J_d}+\frac{\chi}{J_d^2}\right)+p\Omega=\mu_{w}+\hat{\mu}(c_d)+p\Omega\,,
\end{aligned}
\end{equation}
where $\mu_{w}$ is the chemical potential of pure water. Both stress tensor and chemical potential consist of a constitutive component, denoted with an hat in \eqref{oggetti}, and a reactive one. In particular, $\hat{\mu}(c_d)$ (delivering the so-called  \textit{osmotic pressure} $\hat{\mu}(c_d)/\Omega$) and $p\Omega$ are the mixing and mechanical contributions to the chemical potential.

The water flux $\mathbf{h}_d$
\begin{equation}
\label{wf}
\mathbf{h}_d(\mathbf{F}_d,c_d,p)=-\mathbf{M}(\mathbf{F}_d,c_d)\nabla\mu
\end{equation}
depends on the gradient of the chemical potential through the mobility (or, diffusivity) positive definite tensor $\mathbf{M}_d$, that we assume isotropic and linearly dependent on $c_d$,
\begin{equation}
\mathbf{M}(\mathbf{F}_d,c_d)=\frac{D}{\mathcal{R}T}c_d\left(\mathbf{F}_d^T\mathbf{F}_d\right)^{-1}\,,
\end{equation}
where $D$ is the water diffusivity.

We assume that during the whole hydration/dehydration process, the cavity remains filled with water, {\it i.e.}, no cavitation phenomena occur.  Within this hypothesis, at any instant,  the volume $v_\mathcal{C}(t)$ of the cavity coincides with the volume $v_\mathcal{C}^{(w)}(t)$ occupied by the water. This condition is enforced by considering the effective free energy
\begin{equation}
\label{globalpc}
\int_{\mathcal{B}_d}\psi_d^{(r)}\,\text{d}V_d-p_i\left(v_\mathcal{C}-v_\mathcal{C}^{(w)}\right)\,,
\end{equation}
where the unknown Lagrange multiplier $p_i$, represents the pressure field within the cavity.

The balance equations and boundary conditions are obtained from the free energy \eqref{globalpc}, by using the method of virtual powers. Thus, we get 
\begin{equation}
\label{eqgen}
\mathbf{0}=\text{Div}\,\mathbf{S}_d\qquad\text{and}\qquad \frac{\partial c_d}{\partial t}=-\text{Div}\,\mathbf{h}_d\,,
\end{equation}
where, in neglecting the inertial term, we assumed that the solvent diffusion takes place on time scales much longer than any time scales associated with elastic propagation.

By denoting $\muext$ the chemical potential assigned on the external wall, which is assumed to be traction-free, we get the boundary conditions on the outer wall
\begin{equation}
\label{bce}
\mu={\muext}\quad\text{and}\quad\mathbf{S}_d\hat{\mathbf{N}}={\bf 0}\qquad\text{on }\partial_e\mathcal{B}_d\times\mathcal{T}
\end{equation}
and on the inner wall
\begin{equation}
\label{bci}
\mu=\mu_w+p_i\Omega\quad\text{and}\quad\mathbf{S}_d\hat{\mathbf{N}}=-p_iJ_d\mathbf{F}_d^{-T}\hat{\mathbf{N}}\qquad\text{on }\partial_i\mathcal{B}_d\times\mathcal{T}\,.
\end{equation} 
The simulation of air or water exposure of the shell corresponds to considering proper $\muext$ values.

\section{Dehydration-induced circumferential buckling}
At the basis of the instability analysis, there are two key assumptions: (i) before instability occurs, the shell is still axisymmetric and (ii)  the chemo-mechanical state there is determined as quasi-static solution of equations \eqref{eqgen}. These hypotheses allow us to derive the axisymmetric solution of the quasi-static chemo-mechanical problem, as shown in the following. Details regarding calculus in cylindrical coordinates are contained in Appendix \ref{appcalc}.
\subsection{Stress-free swollen solution}
The dry configuration $\mathcal{B}_d$  is a hollow cylinder-shaped shell, of height $L$, with external radius $B$ and thickness $H_d=B-A$, $A$ being the radius of its cavity $\mathcal{C}_d$. Once immersed in water, the thickness of the shell increases until a swollen stress free configuration $\mathcal{B}_o$ is reached. Assuming plain strain deformation, the amount of absorbed solvent is determined by the equation
\begin{equation}
\mu_{w}+\mathcal{R}T\left(\log\frac{J_o-1}{J_o}+\frac{1}{J_o}+\frac{\chi}{J_o^2}\right)+G_d\Omega=\mu_{w}\,,
\end{equation}
that corresponds to assume that the initial state $\mathcal{B}_o$ the shell is a homogeneous state where  $\mu(R)=\mu_{w}$ (chemical equilibrium) and stress tensor $\mathbf{S}_d$ vanishes everywhere. In this in-plane homogeneous configuration, the shell has internal and external radii $A_o$ and $B_o$ equal to $\sqrt{J_o}A$ and $\sqrt{J_o}B$, respectively. We refer to this configuration as the {\it stress-free swollen solution}.

\subsection{Axisymmetric solution}
When the tube is  taken out of the liquid bath and exposed to air, the chemical potential at the  external wall  $\partial_e\mathcal{B}_d$ changes from $\muext=\mu_w$ to $\muext=\mu_a<\mu_w$. Consequently, water begins to be ejected through the outer shell wall.
As long as the water is expelled, the cavity volume reduces and the inner walls $\partial_i\mathcal{B}_d$ undergo an increasing but negative pressure field $p_i$. Such \textit{suction effect} makes the compressive stress at the inner boundary increase and, consequently, the corresponding configuration becomes progressively unstable until buckling occurs, helping the shell to relax the energy stored during the dehydration.


%
Assuming plane strain deformation of the tube cross-section, we first consider purely radial deformations as
\begin{equation}
r=r(R)\,,\qquad\theta=\Theta\,,
\end{equation}
where $(R,\Theta)$ and $(r,\theta)$ are the polar coordinates of a point in the reference dry configuration and in the current one, respectively.
All quantities referred to such axisymmetric configuration are denoted with a subscript \lq$0$\rq. 

The deformation gradient is then $\mathbf{F}_0=\text{diag}\left(r',r/R\right)$, where a prime denotes differentiation with respect to the radial coordinate $R$ 
. Consequently, the local volume constraint \eqref{vc} reduces to
\begin{equation}
\label{01}
r'r=RJ_0\quad\textrm{with}\quad J_0= 1 +\Omega c_0\,, 
\end{equation}
while the Piola-Kirchhoff stress tensor in \eqref{oggetti} becomes
\begin{equation}
\begin{aligned}
\mathbf{S}_0=\text{diag}\left(G_d J_0 Q_0-p_0 Q_0^{-1},G_d Q_0^{-1}-p_0 J_0 Q_0\right)\,,
\end{aligned}
\end{equation}
where  $Q_0\coloneqq R/r$. Furthermore, \eqref{eqgen}$_1$ reduces to
\begin{equation}
\label{02}
p_0'R+G_d\left(J_0Q_0^2-1\right)^2-G_dJ_0'Q_0^2R=0\,,
\end{equation}
where we used the identity $Q_0'=Q_0\left(1-J_0Q_0^2\right)R^{-1}$.

The water flux \eqref{wf} is also assumed purely radial, i.e. $\mathbf{h}_0=\left(h_{0_R},0,0\right)$, with
\begin{equation}
h_{0_R}=-\frac{D}{Q_0^2J_0^2}\left[\frac{-2\chi(J_0-1)+J_0}{\Omega J_0^3}J_0'+\frac{J_0-1}{\mathcal{R}T}p_0'\right]\,,
\end{equation}
and, after a first integration, the quasi-static version of the diffusion equation \eqref{eqgen}$_2$, that is $\text{Div}\,\mathbf{h}_d=0$, yields
\begin{equation}
\label{03}
R\, h_{0_R}=C_0\,,
\end{equation}
where $C_0$ is an integration constant. 

The boundary conditions \eqref{bce} and \eqref{bci} under the quasi-static diffusion assumption also reduce to
\begin{subequations}
\begin{gather}
\mu_0(B)=\mu_a\,,\qquad S_{0_{RR}}(B)=0\,, \label{bc0a} \\ 
\mu_0(A)=\mu_{w}+\Omega p_i\,,\qquad S_{0_{RR}}(A)=-p_i Q_0(A)^{-1}\,, \label{bc0b}
\end{gather}
\end{subequations}
where
\begin{equation}
\label{mu0}
\mu_0=\mu_{w}+\mathcal{R}T\left(\log\frac{J_0-1}{J_0}+\frac{1}{J_0}+\frac{\chi}{J_0^2}\right)+p_0\Omega\,.
\end{equation}
Notice that the two equations in \eqref{bc0b} can be combined to give a unique boundary condition on the inner wall:
\begin{equation}
\label{bc0c}
\mu_0(A)=\mu_{w}-\Omega S_{0_{RR}}(A)Q_0(A)\,.
\end{equation}

 {Finally, in the quasi-static process of cavity emptying, we assume that time variable can be re-parametrized by the radius of the cavity, which is related to the volume of water enclosed. Thus, by assigning the boundary condition} 
\begin{equation}
\label{rA}
r(A)=r_A
\end{equation}
 {and  solving a sequence of equilibrium problems with $r_A$ progressively decreasing,  allows us to emulate the process of cavity emptying.}

\subsection{Incremental analysis}
\label{ia}
As the cavity empties, a critical value of the inner radius (or, equivalently, the enclosed area of the inner wall) is expected at which circumferential buckling occurs.
In order to determine this critical threshold, we introduce a small parameter $\varepsilon$ and the incremental fields $u,v,p_1,J_1$ so that, up to  $O(\varepsilon)$, we have
\begin{align}
\mathbf{x}(R,\Theta,Z)&=\left[r(R)+\varepsilon u(R,\Theta)\right]\hat{\mathbf{e}}_R+\varepsilon v(R,\Theta)\hat{\mathbf{e}}_\Theta\,, \nonumber\\
p(R,\Theta)&=p_0(R)+\varepsilon p_1(R,\Theta) \,,\nonumber\\
J(R,\Theta)&=J_0(R)+\varepsilon J_1(R,\Theta)\,,
\end{align}
with $\{\hat{\mathbf{e}}_R,\hat{\mathbf{e}}_\Theta\}$ as the polar orthonormal basis.
All quantities referred to the incremental configuration are denoted with a subscript \lq$1$\rq.

\subsubsection{Incremental field equations}
The incremental in-plane deformation gradient can be written as
\begin{equation}
\mathbf{F}_1=\left(\begin{matrix}
\partial_R u & R^{-1}\left(\partial_\Theta u -v\right)  \\
\partial_R v & R^{-1}\left(u+\partial_\Theta v\right)  
\end{matrix}\right)\,,
\end{equation}
that, in view of \eqref{vc}, leads to  incremental incompressibility condition $J_0\,\text{Tr}\left(\mathbf{F}_0^{-1}\mathbf{F}_1\right)=J_1$, that can be cast in the form 
\begin{equation}
\label{incr0}
Q_0^{-1}\partial_R u+J_0Q_0R^{-1}\left(u+\partial_\Theta v\right)-J_1=0\,.
\end{equation}
The  incremental Piola-Kirchhoff stress tensor is
\begin{equation}
\mathbf{S}_1=-J_0p_1\mathbf{F}_0^{-T}-J_1p_0\mathbf{F}_0^{-T}+G_d\mathbf{F}_1+J_0p_0\mathbf{F}_0^{-T}\mathbf{F}_1^T\mathbf{F}_0^{-T}    
\end{equation}
and, hence, its components in terms of the unknown fields read as
\begin{align}
S_{1_{RR}}&=p_0 J_0^{-1} Q_0^{-2} \left(\partial_Ru-J_1 Q_0\right)-p_1 Q_0^{-1}+G_d\partial_Ru\,,\nonumber\\
S_{1_{R\Theta}}&=p_0\partial_R v+G_d R^{-1}\left(\partial_\Theta u -v\right)\,,\nonumber\\
S_{1_{\Theta R}}&=p_0 R^{-1}\left(\partial_\Theta u -v\right)+G_d\partial_R v\,,\nonumber\\
S_{1_{\Theta\Theta}}&=-Q_0\left(J_1 p_0+J_0 p_1\right)+R^{-1}\left(G+J_0 p_0 Q_0^2\right)\left(u+\partial_\Theta v\right)\,.
\end{align}
Consequently, equation \eqref{eqgen}$_1$ provides the two linear scalar equations
\begin{align}
\label{incr1}
\partial_R (R S_{1_{RR}})+\partial_\Theta S_{1_{R\Theta}}-S_{1_{\Theta\Theta}}&=0\,,\nonumber\\
\partial_R (R S_{1_{\Theta R}})+\partial_\Theta S_{1_{\Theta\Theta}}+S_{1_{R\Theta}}&=0\,.
\end{align}

In a similar way, we obtain incremental water flux 
\begin{equation}
\mathbf{h}_1=-\mathbf{M}_0\nabla\mu_1-\mathbf{M}_1\nabla\mu_0\,,    
\end{equation}
where $\mu_0$ is given by \eqref{mu0} and
\begin{align}
\mathbf{M}_0&=\frac{D}{\mathcal{R}T}\frac{J_0-1}{\Omega}\left(\mathbf{F}_0^T\mathbf{F}_0\right)^{-1}\,,\nonumber\\
\mathbf{M}_1&=\frac{D}{\mathcal{R}T\Omega}\mathbf{F}_0^{-1}\left[J_1\mathbf{I}-(J_0-1)(\mathbf{F}_1\mathbf{F}_0^{-1}+\mathbf{F}_0^{-T}\mathbf{F}_1^T)\right]\mathbf{F}_0^{-T}\,,\nonumber\\
\mu_1&=-\frac{\mathcal{R}T}{(J_0-1)J_0^3}\left[2\chi(J_0-1)-J_0\right]J_1+\Omega p_1\,.
\end{align}
Thus, the radial and azimuthal components of $\textbf{h}_1$ are
\begin{align}
h_{1_R}&=-\frac{D}{\mathcal{R}T\Omega J_0^3 Q_0^3}\big[\mu_0'Q_0J_0J_1+(J_0-1)\left(-2\mu_0'\partial_Ru+J_0Q_0\partial_R\mu_1\right)\big]\,,\nonumber \\
h_{1_\Theta}&=\frac{D(J_0-1)}{\mathcal{R}T\Omega J_0^2Q_0R}\big[-J_0^2Q_0^3\partial_{\Theta}\mu_1+\mu_0'\left(J_0Q_0^2(\partial_{\Theta}u-v)+R\partial_Rv\right)\big]
\end{align}
respectively, that have to satisfy the incremental diffusion equation 
\begin{equation}
\label{incr2}
\partial_R(Rh_{1_R})+\partial_\Theta h_{1_\Theta}=0\,.
\end{equation}

The local constraint \eqref{incr0} together with  equations \eqref{incr1} and \eqref{incr2} provide a system of four coupled partial differential equations for the unknowns $u,v,p_1,J_1$ in the variable $R$ and $\Theta$, with coefficients depending on the cylindrical solution.
To proceed, similarly to \cite{moulton2011}, we assume the following ansatz for the incremental fields 
\begin{align}
\label{ansatz}
u(R,\Theta)&=\mathcal{U}(R)\cos(n\Theta)\,,\qquad\;\; v(R,\Theta)=\mathcal{V}(R)\sin(n\Theta)\,, \nonumber \\
p_1(R,\Theta)&=\mathcal{P}(R)\cos(n\Theta)\,,\qquad J_1(R,\Theta)=\mathcal{J}(R)\cos(n\Theta)\,,
\end{align}
with $n$ an integer representing the buckling mode, {\it i.e.} the number of folds in the buckled state. In this way, the incremental problem simplifies  to a system of ordinary differential equations.

From the equation \eqref{incr0} we obtain
\begin{equation}
\label{J}
\mathcal{J}=Q_0^{-1}\mathcal{U}'+J_0Q_0R^{-1}\left(\mathcal{U}+n\mathcal{V}\right)\,,
\end{equation}
that allows to eliminate $\mathcal{J}$ in the differential equations. Furthermore, to deal with \eqref{incr2} it is convenient, from a computational point of view, to consider the following expansion for the radial component of the incremental flux
\begin{equation}
h_{1_R}(R,\Theta)=\mathcal{H}(R)\cos(n\Theta)
\end{equation}
and consider its amplitude $\mathcal{H}$ as a further unknown. In so doing,  by introducing the vector of unknowns $\mathbf{q}\coloneqq\left(\mathcal{U},\mathcal{U}',\mathcal{V},\mathcal{V}',\mathcal{P},\mathcal{H}\right)$, the system of ODEs can be recast in the form
\begin{equation}
\label{increq}
\mathbf{q}'=\mathbf{A}_n\left(R,Q_0(R),p_0(R),J_0(R)\right)\mathbf{q}\,,
\end{equation}
where $\mathbf{A}_n$ is the $6\times 6$ coefficient matrix, whose expressions are listed in Appendix \ref{appcoeff}.\\
\subsubsection{Incremental boundary conditions}
Before writing the incremental boundary conditions, let's notice that the first order term of the cavity volume $v_\mathcal{C}$ vanishes. Indeed,
\begin{equation}
\label{vol}
v_\mathcal{C}=\int_{\mathcal{C}_d}J_d\text{d}V=L\pi r_A^2+\varepsilon \int_{\mathcal{C}_d}J_1\text{d}V
\end{equation}
and the integral on the right hand side of \eqref{vol} is zero, due to the ansatz \eqref{ansatz}. As a consequence, the unknown pressure field $p_i$, that is the Lagrangian multiplier enforcing the inner volume constraint, also remains unchanged up to the first order. 

Then, the six boundary conditions, obtained by expanding \eqref{bce} and \eqref{bci} to the order $\varepsilon$, read as
\begin{align}
\label{bc1}
\mu_1(B)&=0\,,\qquad \mathcal{S}_{1_{RR}}(B)=0\,,\qquad \mathcal{S}_{1_{\Theta R}}(B)=0\,, \nonumber \\
\mu_1(A)&=0\,, \qquad
\mathcal{S}_{1_{\Theta R}}(A)=-p_i\Big(\mathcal{V}(A)+n\mathcal{U}(A)\Big)R^{-1}\,, \nonumber \\
\mathcal{S}_{1_{RR}}(A)&=-p_i\Big(\mathcal{J}(A)Q_0(A)-\mathcal{U}'(A)\Big)J_0(A)^{-1}Q_0(A)^{-2}\,,
\end{align}
where $p_i$ and $\mathcal{J}$ are given by \eqref{bc0b}$_2$ and \eqref{J}, respectively, while $\mathcal{S}_{1_{RR}}$ and $\mathcal{S}_{1_{\Theta R}}$ are the amplitudes of the incremental stress tensor components $S_{1_{RR}}=\mathcal{S}_{1_{RR}}(R)\cos(n\Theta)$ and $S_{1_{\Theta R}}=\mathcal{S}_{1_{\Theta R}}(R)\sin(n\Theta)$.

\subsubsection{Resuming the incremental boundary value problem}
The goal is to find a value $r_A$ of the cavity radius  for which the incremental system \eqref{increq} of ordinary differential equations admits a non-axisymmetric solution. We find convenient  to introduce,  in lieu of $r_A$, the  dimensionless parameter
\begin{equation}
\beta\coloneqq \frac{v_\mathcal{C}}{v_{\mathcal{C}_o}}
\end{equation} 
representing the ratio between the actual volume of the cavity $v_\mathcal{C}=L\pi r_A^2$ and the swollen one $v_{\mathcal{C}_o}=L\pi r_{A_o}^2$. We expect to find a critical value $\betac<1$ at which a transition from the  axisymmetric to a wrinkled shape occurs, as the former becomes unstable.

\begin{figure}[t]
\centering
\includegraphics[scale=1.6]{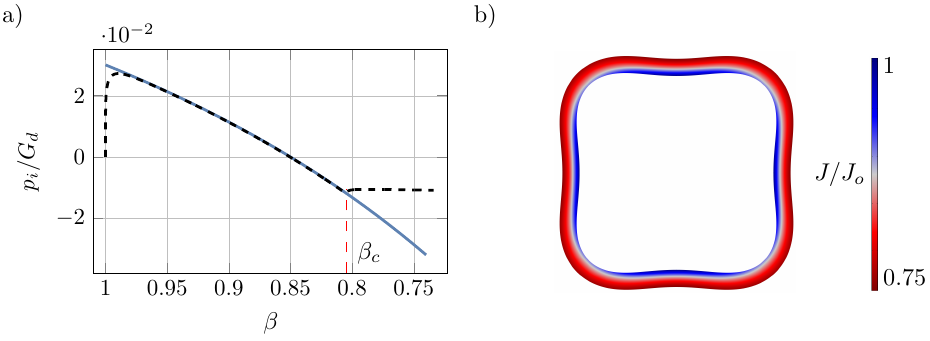}
\caption{Dimensionless cavity pressure $p_i/G_d$ as a function of the dimensionless enclosed volume $\beta$, comparison between quasi-static model (solid curve) and FEM simulation (dashed curve).  Cross section shape of the shell at the critical point (b).}
\label{pibeta}
\end{figure}

As the coefficients of the incremental equations are determined by the solution of the $O(1)$ problem and the parameter $\beta$ only appears in the boundary condition \eqref{rA} of the $O(1)$ problem, we need to solve the zeroth-order equations \eqref{01}, \eqref{02}, \eqref{03} and the first-order equations \eqref{increq} simultaneously. For this purpose, the problem can be written as a unique system of first order differential equations of the form
\begin{equation}
\mathbf{y}'=\mathbf{f}\left(\mathbf{y},R\right)\,,
\end{equation}
where $\mathbf{y}\coloneqq\left(r,p_0,J_0,\mathcal{U},\mathcal{U}',\mathcal{V},\mathcal{V}',\mathcal{P},\mathcal{H}\right)$ is the vector of unknowns, to which we must add the two unknown constants $C_0$ and $\beta$. Finally, we need a total of eleven boundary conditions: ten of these are given by equations \eqref{bc0a}, \eqref{bc0c}, \eqref{rA} and \eqref{bc1}, while the eleventh is $\mathcal{U}(A)\neq 0$ and imposes a nontrivial solution of the problem.

\section{Results and discussion}
Solutions of the above boundary value problem are obtained by using \texttt{bvp4c} in \texttt{Matlab}, with constitutive and geometric parameters  chosen as in Table \ref{tab}.
\begin{table}[h]
\begin{center}
\caption{Values of parameters used in numerical integration.}
\hspace{1cm}
\begin{tabular}{ c c c c c c c }
\toprule
$G_d$ (Pa) & $\chi$ & $\Omega\,(\text{m}^3\text{mol}^{-1})$ & $D\,(\text{m}^2\text{s}^{-1})$ & $T$ (K) & $B$ (m) & $A$ (m)\\
\midrule
$5\cdot 10^7$ & $0.2$ & $1.8\cdot 10^{-5}$ & $10^{-9}$ & $293$ & $10^{-2}$ & $B-1.25\cdot 10^{-3}$ \\
\bottomrule
\end{tabular}
\label{tab}
\end{center}
\end{table}

%
These solutions are then compared with the FEM simulations of the fully dynamic problem. Thus, equations (\ref{eqgen}) together with constitutive equations (\ref{oggetti}, \ref{wf}) and constraints (\ref{vc}), (\ref{globalpc}), (\ref{bce}), (\ref{bci}) are rewritten in weak formulation and the full problem is restated as follows: find the value of the variables $\mathbf{u}_d$, $c_d$, $p$, $p_i$ and $c_s$ (an auxiliary concentration variable used in the chemical boundary conditions) such that, for any test functions $\Tilde{\mathbf{u}}_d$, $\Tilde{c}_d$, $\Tilde{p}$, $\Tilde{p}_i$ and $\Tilde{c}_s$, balance
equations (\ref{eqgen}), volumetric constraints (\ref{vc}), (\ref{globalpc}), boundary conditions (\ref{bce}), (\ref{bci}) and initial conditions hold. The tube cross-section under plane deformation is discretized by a mapped mesh yielding about 20 thousands degrees of freedom. The problem is then solved in time using all quadratic Lagrange form functions, except for the variable $p$, for which a discontinuous linear Lagrange form function is used.

We first focus on the solution before the buckling occurs. In order to simulate the transition from water bath to air exposure of the tube, an external chemical potential is assigned in the FEM simulation that varies rapidly over time from the initial water potential to the plateau potential of the air.
Figure \ref{pibeta}a shows how the dimensionless pressure $p_i/G_d$ on the cavity walls changes as dehydration proceeds. In particular, our quasi static analysis (blue line) is here compared with what is obtained from FEM simulation of the dynamic process (dashed black line). The two curves show a remarkable agreement where the external chemical potential has reached its plateau value $\mu_a$, that is equal to $-2\cdot 10^3$ J/mol once $\mu_w$ is taken as reference.

 {We stress that, while in the dynamic problem the applied chemical potential changes over time with an increasing ramp until reaching a plateau value, in the quasi-static problem, the assigned chemical potential correspond to the plateau value (the air chemical potential). This is why the two curves differ so much on the left-hand side of the Figure \ref{pibeta}. On the other hand, when in the dynamic problem the external potential reaches the plateau value, the two curves are in fact indistinguishable until the buckling occurs.}

\begin{figure}[t]
\centering
\includegraphics[scale=0.35]{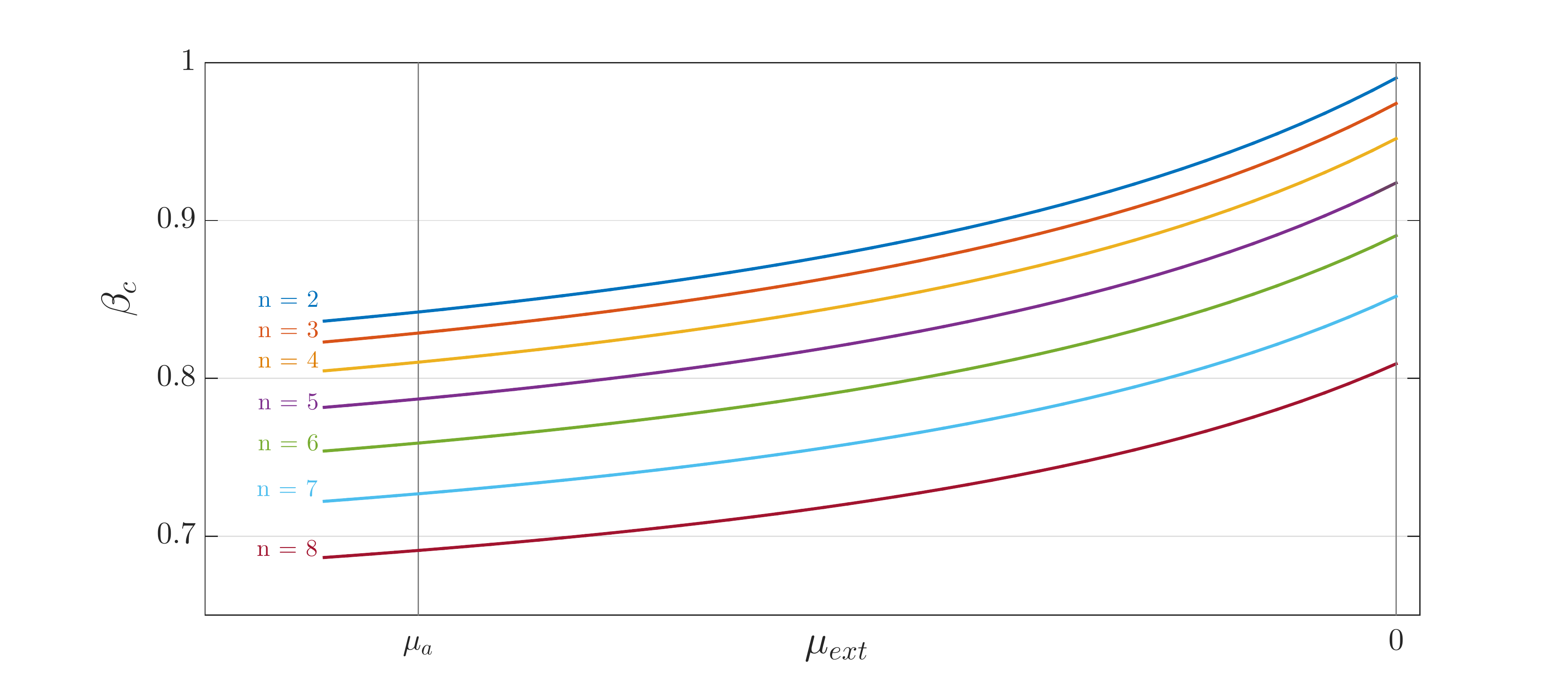}
\caption{Critical ratio $\betac$ as a function of the external chemical potential $\mu_e$,  for modes $n=2,..,8$. The vertical black line corresponds to the air chemical potential $\mu_a = -2\cdot 10^3$ J/mol.}
\label{betami}
\end{figure}

As diffusion goes on, the inner pressure becomes negative, effectively realising a {\it suction effect} on the cavity walls, which is responsible for the circumferential buckling. In accordance with FEM simulation,  at   $\beta\approx 0.806$  the tube section buckles into four-sided  shape, as depicted in Figure \ref{pibeta}b. On the other hand, the bifurcation analysis yields $\beta_{cr}=0.809$, for the buckling mode $n=4$, which is in good agreement with the FEM simulation.

However, the perturbation analysis provides a critical threshold for each mode of buckling, for any applied chemical potential. Figure \ref{betami}, sketches  $\betac$  as a function of the applied chemical potential $\muext$, for various modes $n$. In particular, the vertical black line corresponds to the air chemical potential $\mu_a = -2\cdot 10^3$ J/mol. We note that, for any fixed value of $\muext$, in the emptying process, {\it i.e.} gradually decreasing $\beta$,  the first critical threshold encountered is that of the mode $n=2$. Furthermore, the critical thresholds are decreasing functions of the respective buckling mode $n$, which means that the higher modes are activated later. 

By virtue of the above, we should expect that, in the emptying phase, the axisymmetric solution bifurcates into the $n=2$ mode, instead of the $n=4$ mode predicted by the FEM. The reason for this contradiction could be the subject of future work. Here we simply point out that, for the same buckling mode, bifurcation analysis and the FEM simulation give consistent results.

\begin{figure}[t]
\centering
\includegraphics[scale=0.45]{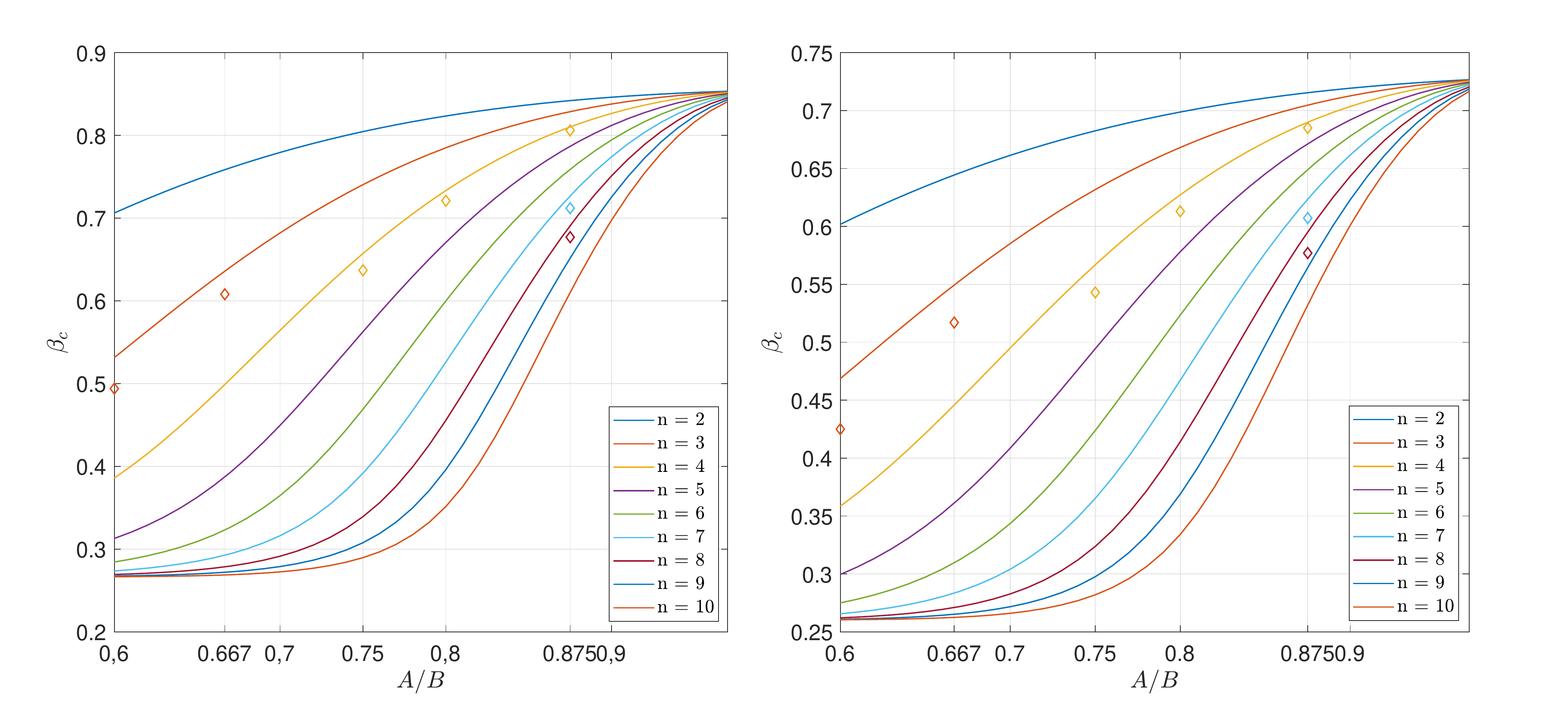}
\caption{Critical ratio $\betac$ as function of the shell thickness $A/B$, with $m=0.37$ (left) and $m=0.16$ (right). It is worth noting that in \cite{curatolo2021} it was observed an intersection between the different curves for the spherical shell while here in the case of the cylinder the colored curves do not intersect each other. }
\label{betarat}
\end{figure}

 {Another effect worth investigating through bifurcation analysis, in analogy to other work \cite{moulton2011, curatolo2021}, is the influence of relative shell thickness on critical threshold.} Figure \ref{betarat} shows the critical threshold as a function of the dimensionless thickness $A/B$, when $\muext = \mu_{\rm a}$. The two graphs in Figure \ref{betarat} correspond to two different values of the dimensionless parameter
$m\coloneqq G_d\Omega/(\mathcal{R}T)$, which measures the hydrogel ability  to be permeated by the solvent.
Note that for each $A/B$ value, the modes preserve the order of activation, i.e. the more wrinkled solutions  are triggered with successively smaller critical thresholds. This feature is not shared by spherical  shells \cite{curatolo2021}, where thinner shells bifurcate in higher modes shapes and where critical thresholds are not monotonic functions of $n$. Moreover, as $m$ decreases all thresholds are lowered, which means that to make a very permeable shell unstable, more cavity emptying is required.

\begin{figure}[t]
\centering
\includegraphics[scale=2]{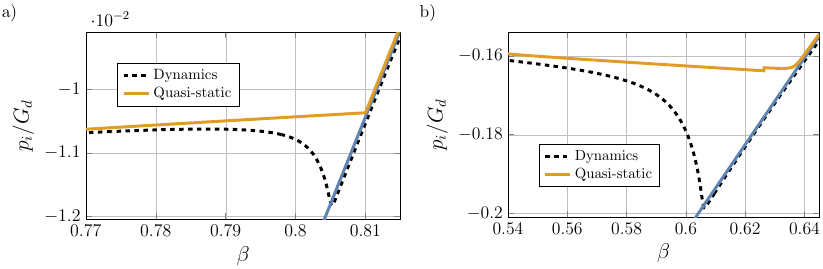}
\caption{Dimensionless cavity pressure as a function of $\beta$. The solid blue line refers to quasi-static axisymmetric solution, while the orange line corresponds to FEM simulation of the static model. The dashed black line refers to the FEM of the fully dynamic model. The values of the parameters used for the simulations are those in Table 1, choosing $A/B=0.875$ ($n=4$) for the left-hand graph and $A/B=0.667$ ($n=3$) for the right-hand graph. }
\label{fig5}
\end{figure}

 {The diamond-shape markers in Figure \ref{betarat} correspond to critical threshold   obtained by FEM simulation. The agreement with the results provided by the bifurcation analysis is good for thin shells, while it seems to deteriorate (although still satisfactory) for thick cells. We think this difference is due to the rough approximation of the dynamic model with the quasi-static model used for the incremental analysis, which is not appropriate in the vicinity of the critical threshold where there is a sudden change in shape.  To illustrate this, a comparison of the FEM simulations of the dynamic model and the corresponding quasi-static model is shown in the Figure \ref{fig5}. Two aspects emerge: the quasi-static model simulation, which provides a transcritical phase transition, predicts a critical threshold in agreement with the bifurcation analysis.  In contrast, the transition predicted by the full model is slightly delayed compared to the expected threshold. This delay is all the greater, the thicker the threshold.  However, in the phases following the instability, the dynamic solution relaxes towards the quasi-static solution.}

\section{Conclusions}
We studied the chemomechanical effect of dehydration on the circumferential instability of an elastic cylindrical tube. During dehydration, the tube drains water through the walls, deflating and emptying the cavity. This triggers a suction effect on the inner wall that causes the circumferential instability of the tube. 

The tube is modelled as a Neo-Hookean elastic, incompressible solid, that can inflate/deflate by absorbing/expelling water. The diffusion of water in the body is governed by the Flory-Rehener equation.  Buckling is examined as a bifurcation problem using the formalism of incremental equations, under the assumption of a quasi-static process. The analysis captures some relevant aspects of instability by predicting different modes of bifurcations, with more or less wrinkled profiles, and their critical thresholds. However, chemo-mechanical coupling complicates the perturbation analysis compared to the analogous problem for pressurised or incrementally growing shells (see \cite{moulton2011} and references therein), as the pressure causing the instability is not a directly controllable parameter but is instead an unknown quantity.  

In parallel, we simulated, via FEM based on the full dynamic model, the emptying of a hydrogel tube from a water bath to being exposed with the outer wall to air. The results found with the two methods are consistent, and are in excellent agreement in the limit of thin shells.

\section*{Acknowledgments}
This manuscript was also conducted under the auspices of the GNFM-INdAM.
\section*{Funding}
The work of G.N. has been funded by the MIUR  Project PRIN 2020, “Mathematics for Industry 4.0”, Project No. 2020F3NCPX. The work of F.L. has been funded by POR Puglia FESR FSE 2014-2020. 
%
%

\appendix
\section{Calculus in cylindrical coordinates}
\label{appcalc}
Given a scalar-valued field  in cylindrical coordinates $f(R,\Theta,Z)$, the gradient of $f$ is
\begin{equation}
\nabla f = \partial_R f\, \hat{\mathbf{e}}_R+R^{-1}\partial_\Theta f\, \hat{\mathbf{e}}_\Theta+\partial_Z f\, \hat{\mathbf{e}}_Z\,,
\end{equation}
where $\left\lbrace\hat{\mathbf{e}}_R,\hat{\mathbf{e}}_\Theta,\hat{\mathbf{e}}_Z\right\rbrace$ is the standard orthonormal basis in cylindrical coordinates.

Let $\mathbf{v}= v_R(R,\Theta,Z) \ev_R + v_\Theta(R,\Theta,Z) \ev_\Theta +v_Z(R,\Theta,Z) \ev_Z$ then
\begin{align}
\nabla\textbf{v}=\;
& \partial_R v_R \; \ev_R \otimes \ev_R 
+ R^{-1}  \left(\partial_\Theta v_R - V_\Theta\right)  \; \ev_R \otimes \ev_\Theta 
+ \partial_Z v_R   \; \ev_R \otimes \ev_Z \nonumber \\
+\; & \partial_R v_\Theta\;   \ev_\Theta \otimes \ev_R 
+ R^{-1}\left(\partial_\Theta v_\Theta +v_R\right)  \; \ev_\Theta \otimes \ev_\Theta 
+ \partial_Z v_\Theta  \;  \ev_\Theta \otimes \ev_Z \nonumber \\
+\; &\partial_R v_Z \; \ev_Z \otimes \ev_R
+ R^{-1}\partial_\Theta v_Z  \; \ev_Z \otimes \ev_\Theta 
+ \partial_Z v_Z  \; \ev_Z \otimes \ev_Z \,,
\end{align}
and, hence, 
\begin{equation}
\text{Div}\,\mathbf{v} = R^{-1}\left[\partial_R(R v_R)+\partial_\Theta v_\Theta\right]+\partial_Z v_Z \,.
\end{equation}

Let $\mathbf{S}$ be a tensor  then the components of the divergence of $\mathbf{S}$ are
\begin{align}
[\text{Div} \; \mathbf{S}]_R & = R^{-1}{\partial_R (R S_{RR})}+R^{-1} {\partial_\Theta S_{R\Theta}}-R^{-1} {S_{\Theta\Theta}}+{\partial_Z S_{RZ}},\nonumber\\
[\text{Div} \; \mathbf{S}]_\Theta & = R^{-1} \partial_R (R S_{\Theta R}) +R^{-1}\partial_\Theta S_{\Theta \Theta}  +R^{-1} S_{R\Theta} + \partial_Z S_{\Theta Z},\nonumber\\
[\text{Div}\;\mathbf{S}]_Z & = R^{-1}{\partial_R (R S_{Z R})} +R^{-1}{\partial_\Theta S_{Z\Theta}}+{\partial_Z S_{Z Z}}\,.
\end{align}

\section{Coefficients of the ODE system}
\label{appcoeff}
Let
\begin{equation}
    \mathcal{G}_0\coloneqq\frac{\text{d}\mu_{0}}{\text{d}J_0}\,.
\end{equation}
The non-vanishing coefficients of the linear system \eqref{increq} are
\begin{align}
    A_{12}&=1\,,\nonumber\\
    A_{21}&=\left[R^2 \left(J_0-1\right) \left(G_d \Omega  Q_0^2+\mathcal{G}_0\right)\right]^{-1}Q_0^2(J_0-1)\,\times\nonumber\\&\quad\times\left\{G_d\Omega(1+n^2)+J_0^2Q_0^2\mathcal{G}_0-R\left[(J_0\mathcal{G}_0)'-\Omega p_0'+\mu_0'J_0(J_0-1)^{-1}\right]\right\}\,,\nonumber \\
    A_{22}&=\frac{-J_0(J_0-1)\left[G_d\Omega Q_0^2+R\mathcal{G}_0'+(2J_0 Q_0^2-1)\mathcal{G}_0-\mu_0'R J_0^{-1}\right]-\mu_0'R}{R J_0\left(J_0-1\right) \left(G_d \Omega  Q_0^2+\mathcal{G}_0\right)}\, ,\nonumber\\
    A_{23}&=\frac{n Q_0^2 \left\{(J_0-1)\left[2G_d\Omega-R(J_0\mathcal{G}_0)'+R\Omega p_0'+J_0^2Q_0^2\mathcal{G}_0\right]-\mu_0'RJ_0\right\}}{R^2 \left(J_0-1\right) \left(G_d \Omega  Q_0^2+\mathcal{G}_0\right)}\,,\nonumber \\
    A_{24}&=-\frac{n J_0 Q_0^2 \mathcal{G}_0}{R(G_d\Omega  Q_0^2+ \mathcal{G}_0)}\,,\qquad\qquad A_{26}=-\frac{\mathcal{R}T\Omega J_0^2 Q_0^3}{D\left(J_0-1\right) \left(G_d \Omega  Q_0^2+\mathcal{G}_0\right)}\,,\nonumber \\
    A_{34}&=1 \,,\nonumber\\
    A_{41}&=\frac{n \left(2 G_d+R p_0'\right)}{G_d R^2}, \qquad\qquad A_{43}=\frac{(n^2+1)G_d+R p_0'}{G_d R^2}\,,\nonumber\\
    A_{44}&=-\frac{1}{R}\,, \qquad\qquad A_{45}=-\frac{n J_0 Q_0}{G_d R}\,,\nonumber\\
    A_{51}&=-\left[R^2 \left(J_0-1\right) \left(G_d \Omega  Q_0^2+\mathcal{G}_0\right)\right]^{-1}Q_0\,\times\big\{\Big. G_d\mu_0'RJ_0Q_0^2\,+\nonumber\\&\quad +(J_0-1)\left[R(\mathcal{G}_0p_0'+G_dQ_0^2(J_0\mathcal{G}_0)')+G_d\mathcal{G}_0(1+n^2-J_0^2Q_0^4)\right]\Big.\big\}\,,\nonumber \\
    A_{52}&=\frac{G_d Q_0 \left\{(J_0-1)\left[R(\mu'_0-J_0\mathcal{G}'_0)+2J_0\mathcal{G}_0(1-J_0Q_0^2)\right]-\mu'_0R\right\}}{R J_0\left(J_0-1\right) \left(G_d \Omega  Q_0^2+\mathcal{G}_0\right)}\,,\nonumber \\
    A_{53}&=-\left[R^2 \left(J_0-1\right) \left(G_d \Omega  Q_0^2+\mathcal{G}_0\right)\right]^{-1}n Q_0\,\times\big\{\Big.G_d\mu'_0RJ_0Q_0^2\,+\nonumber\\&\quad+\,(J_0-1)\left[R(\mathcal{G}_0p_0'+G_dQ_0^2(J_0\mathcal{G}_0)')+G_d\mathcal{G}_0(2-J_0Q_0^4)\right]\big.\big\} \,,\nonumber\\
    A_{54}&=-\frac{G_dnJ_0 Q_0^3 \mathcal{G}_0}{R\left(G_d \Omega  Q_0^2+\mathcal{G}_0\right)}\,, \qquad\qquad A_{56}=-\frac{G_d \mathcal{R} T\Omega J_0^2 Q_0^4}{D \left(J_0-1\right) \left(G_d \Omega  Q_0^2+\mathcal{G}_0\right)} \,,\nonumber\\
    A_{61}&=\frac{Dn^{2}(J_{0}-1)Q_0\left(R\mu_{0}'-J_{0}^{2}Q_{0}^{2}\mathcal{G}_{0}\right)}{\mathcal{R}T\Omega J_{0}R^{3}}\,,\qquad\qquad A_{62}=-\frac{Dn^{2}(J_{0}-1)Q_{0}\mathcal{G}_{0}}{\mathcal{R}T\Omega R^{2}}\,,\nonumber\\
    A_{63}&=\frac{Dn(J_{0}-1)Q_0\left(R\mu_{0}'-J_0^2 Q_0^2 n^2\mathcal{G}_{0}\right)}{\mathcal{R}T\Omega J_{0}R^{3}}\,,\qquad\qquad A_{64}=-\frac{Dn(J_{0}-1)\mu_0'}{\mathcal{R}T\Omega J_{0}^2 Q_0 R}\,,\nonumber\\
    A_{65}&=-\frac{Dn^{2}(J_{0}-1)Q_0^2}{\mathcal{R}T R^{2}}\,,\qquad\qquad A_{66}=-\frac{1}{R}\,.
\end{align}





\end{document}